\documentclass[aps,prl,twocolumn,superscriptaddress,noeprint]{revtex4-2}

\usepackage{natbib}
\usepackage{graphicx}
\usepackage{xcolor}
\usepackage{rotating}
\usepackage{bm,amsmath,amssymb}
\usepackage[utf8]{inputenc}
\usepackage{amsmath,amsthm}

\usepackage[colorlinks=true, urlcolor=blue, linkcolor=blue, citecolor=blue, hyperindex=true, linktocpage=true]{hyperref}

\usepackage{cleveref}

\crefname{Ass}{Assumption}{Assumptions}

\begin{document}

\title{Bulk Spacetime Encoding via Boundary Ambiguities}

\author{Zhenkang Lu}
\email{zhenkanglu9@gmail.com}
\affiliation{%
 Department of Physics, Shanghai University, Shanghai, 200444, China
 }%

\author{Cheng Ran}
\email{r\_cheng@shu.edu.cn}
\affiliation{%
 Department of Physics, Shanghai University, Shanghai, 200444, China
 }%

\author{Shao-Feng Wu}
\email{sfwu@shu.edu.cn}
\affiliation{%
 Department of Physics, Shanghai University, Shanghai, 200444, China
 }%
\affiliation{Center for Gravitation and Cosmology, Yangzhou University, Yangzhou, 225009, China}

\begin{abstract}
\noindent
We propose a method to reconstruct the metric and its arbitrary-order derivatives at the horizon for any static, planar-symmetric black hole, using an infinite set of discrete pole-skipping points in momentum space where the boundary Green's function becomes ambiguous. This method is fully analytical and involves solving only linear equations. The near-horizon reconstruction can extend either inside or outside the horizon until reaching the nearest singularity in the complex radial plane. It further enables a reinterpretation of any pure gravitational field equation in pole-skipping data. Moreover, our method reveals that the pole-skipping points are redundant: only a subset is independent, while the rest are fixed by an equal number of homogeneous polynomial constraints. These identities are universal, independent of the details of the bulk geometry, including its dimensionality, asymptotic behavior, or the existence of a holographic duality.
\end{abstract}

\maketitle

{\it Introduction---}A central question in holographic duality \cite{witten_1998_Holography_2, Gubser_1998_holography, maldacena_1999_first_Holography} is how a smooth, classical spacetime described by Einstein’s equations, emerges from the intricate quantum data of a boundary theory. While quantum entanglement is often regarded as the ``thread'' weaving spacetime together \cite{maldacena_2003_extend_AdS_and_Two_entangled_CFT, ryu_2006_rt_formula_original, raamsdonk_2010_build_up_Spacetime_entanglement, swingle_2012_entanglement_renormalization_geometry, maldacena_2013_ER_equal_to_EPR}, the precise mechanism remains elusive.

One difficulty stems from the fact that entanglement is hard to compute and measure in quantum many-body systems. More fundamentally, entanglement alone is insufficient to capture the entire spacetime \cite{susskind_2014_entanglement_not_enough, Susskind_2014_CV_original}. Indeed, among the numerous studies exploring the spacetime reconstruction from dual field theories \cite{deharo_2001_other_T_munv_Metric_reconstruction, hammersley_2006_Bulk_reconstruction_null_geodesic, bilson_2008_Correlation_function_bulk_metric_1, bilson_2011_Entanglement_reconstruct_bulk_metric_2, czech_2014_Entanglement_reconstruct_bulk_metric_3, engelhardt_2017_Other_light_cone_Metric_reconstruction_2, roy_2018_entanglement_modularHamiltonian_bulk_metric_reconstruction, Hashimoto_2018_AdS/DL_original, bao_2019_Entanglement_reconstruct_bulk_metric_1, yan_2020_Metric_reconstruction_shear_viscosity, hashimoto_2021_Wilson_loops_bulk_reconstruction_metric_1, hashimoto_2021_complexity_bulk_metric_1, xu_2023_Entanglement_reconstruct_bulk_metric_ML_1, yang_2023_Greenfunction_to_metric_analytical}, the only established method for encoding the interior geometry of black holes relies on the `complexity=volume' conjecture \cite{hashimoto_2021_complexity_bulk_metric_1, xu_2023_Entanglement_reconstruct_bulk_metric_ML_1}.

These challenges motivate us to pursue an approach capable of reconstructing both the interior and exterior of black holes. Ideally, it should be analytical, simple and transparent, free from auxiliary conjectures, and testable through experiments or numerical simulations.

In this Letter, we present an approach that does not rely on entanglement and complexity. We demonstrate that the entire black hole spacetime can be encoded through characteristic structures in linear response data. The key element is the phenomenon of pole-skipping, where poles and zeros of holographic Green’s functions collide at specific complex frequencies and momenta, rendering the correlator ambiguous \cite{grozdanov_2018_upper_Chaos_pole_skipping_0, blake_2018_upper_Chaos_pole_skipping_1, blake_2018_Upper_Chaos_pole_skipping_2}. While pole-skipping is best known for its connection to quantum chaos \cite{grozdanov_2018_upper_Chaos_pole_skipping_0, blake_2018_upper_Chaos_pole_skipping_1, blake_2018_Upper_Chaos_pole_skipping_2, Haehl_2018_EFT_chaotic_CFT, Grozdanov_2019_higher_derivative_correction, Liu_2020_TMG_chaos_pole-skipping, grozdanov_2021_univalence_bound_pole-skipping, ramirez_2021_Pole_skipping_chaos_CFT_2, Blake_2021_pole-skipping_rotating_black_hole, baggioli_2022_univalence_axion_models, knysh_2024_Horizon_symmetry_pole, chua_2025_Replica_pole_skipping},  our work instead exploits a distinct infinite tower of pole-skipping points that resides in the lower-half frequency plane \cite{grozdanov_2019_lower_infinite_pole_skipping_2, natsuume_2019_Gauss-Bonnet_correction_pole-skipping_2, wu_2019_gauss-Bonnet_correction_pole-skipping_k_change_w_unchange, blake_2020_lower_infinite_pole_skipping_1, ceplak_2020_Pole_Skipping_fermion_BTZ_analytical, choi_2021_Pole_skipping_SYK_chain, ahn_2021_Classify_pole_skipping, natsuume_2021_Pole_skipping_zero_temperature, yuan_2021_near-Horizon-analysis-Lif-Rindler-AdS, abbasi_2021_Pole_skipping_chaos_bound_QNM, grozdanov_2023_Pole_skipping_hyperbolic_sphereical_flat, Natsuume_2023_pole-skipping_non-black_hole, grozdanov_2024_Constraints_on_spectrum_infinte_product, Grozdanov_2025_4d_spectral_constraint_2}.

Our reconstruction applies to general static, planar-symmetric black holes in arbitrary dimensions.  Remarkably, it is fully analytical and requires solving nothing more than linear equations. This explicit approach extends naturally to the vacuum Einstein equation, with any higher curvature correction, allowing its complete reformulation in boundary pole-skipping data.

One of the most striking outcomes of our approach is that the boundary ambiguities encoding the geometry are highly redundant. This is reminiscent of the redundant encoding in holographic quantum error correction, where the deep bulk information is protected against local boundary erasures \cite{Almheiri_2015_Error_correcting_code_original, Pastawski_2015_Happy_code}. In our framework, such redundancy manifests as an infinite set of homogeneous polynomial identities constraining the pole-skipping momenta. Crucially, these identities are universal, independent of the specifics of the bulk theory and even of holography itself.

With these features combined, our approach motivates experimental searches for pole-skipping, tests of its redundancy, and offers new strategies to probe emergent geometry in diverse quantum systems such as strange metals and the quark–gluon plasma. This Letter provides a concise summary of our companion paper \cite{Lu_2024_Longer_paper}.

{\it Near-horizon analysis---}As the critical ingredient in our reconstruction, the location of boundary pole-skipping points can be readily identified from the bulk equation of motion using a near-horizon analysis \cite{blake_2018_Upper_Chaos_pole_skipping_2, blake_2020_lower_infinite_pole_skipping_1}. We briefly review this procedure to set the stage for our reconstruction method \footnote{Alternatively, pole-skipping points can be obtained through the covariant expansion formalism \cite{wang_2022_Pole-skipping_gauge_bosonicfields, ning_2023_pole-skipping_gauge_fermionicfields}.}.

Consider a static, planar-symmetric black hole in $d+2$ dimensions. In ingoing Eddington–Finkelstein coordinates, the metric takes the form
\begin{equation}
\label{equ_general_metric_d}
ds^2 = -g_{vv}(r)dv^2 + 2g_{vr}(r) dvdr + r^2 d\vec{x}^2,
\end{equation}
where $\vec{x}$ spans the $d$-dimensional spatial directions. In a holographic context, a thermal CFT with temperature $T_b$ exists at the boundary $r \to \infty$.

Now, take a massless probe scalar field $\phi$ coupled to the background, which satisfies the Klein-Gordon (KG) equation $\nabla^{2}\phi=0$. This scalar field is dual to a boundary marginal operator $\mathcal{O}$ with scaling dimension $\Delta=d+1$. After performing a Fourier transform $\phi = \varphi(r)e^{-i\omega v+i\vec{k}\vec{x}}$, the KG equation becomes
\begin{widetext}
    \begin{equation}\label{equ_KGequation_General_d} 
    \frac{g_{vv}(r)}{g_{vr}(r)^2}\varphi^{\prime\prime}(r)+\left (-\frac{2 i\omega}{g_{vr}(r)} + \frac{d g_{vv}(r)}{r g_{vr}(r)^2} - \frac{g_{vv}(r) g_{vr}^{\prime}(r)}{g_{vr}(r)^3} + \frac{g_{vv}^{\prime}(r)}{g_{vr}(r)^2}\right)\varphi^{\prime}(r)-\left(\frac{\mu}{r^{2}}+\frac{i d\omega}{r g_{vr}(r)}\right)\varphi (r)=0, 
    \end{equation}
\end{widetext}
where we have defined $\mu = k^2$. Without loss of generality, we set $r_h = 1$ in the following. The metric components admit the near-horizon expansions:
\begin{equation}\label{equ_gvvgvr_expansion}
\begin{aligned}
& g_{vv}(r)=g_{vv_1}(r-1)+g_{vv_2}(r-1)^2+\ldots \\
& g_{vr}(r)=g_{vr_0}+g_{vr_1}(r-1)+g_{vr_2}(r-1)^2+\ldots \\
\end{aligned}
\end{equation}
In this setup, the Hawking temperature is given by \mbox{$T_h = \frac{g_{vv_1}}{4\pi g_{vr_0}}$}. Meanwhile, the field $\varphi(r)$ admits a near-horizon expansion of the form
\begin{equation}\label{equ_Taylor_Expansion_rh} 
\varphi (r) = (r-1)^\alpha \sum_{p=0}^\infty \varphi_p (r-1)^p, 
\end{equation} 
with two exponents: $\alpha_1 = 0$ and $\alpha_2 = \frac{i\omega}{2\pi T_{h}}$. Typically, for general $\omega$, only the $\alpha_1 = 0$ solution is ingoing at the horizon. Under this condition, the boundary retarded Green’s function of $\mathcal{O}$, denoted $\mathcal{G}^{\mathcal{O}}_{R}(\omega, \mu)$, can be uniquely determined as the ratio of the normalizable to the non-normalizable mode. However, at specific frequencies $\omega_n = -i 2\pi n T_{h}$ for $n \in \mathbb{Z}^+$ and certain values of $\mu_n$, both modes become regular at the horizon, rendering ambiguous $\mathcal{G}^{\mathcal{O}}_R(\omega_{n}, \mu_{n})$.

These special points are identified through performing near-horizon analysis. To begin with, we expand $\varphi(r)$, $g_{vv}(r)$ and $g_{vr}(r)$ around the horizon up to $n^{th}$ order and substitute them into the KG equation \eqref{equ_KGequation_General_d}. By requiring the coefficients of $(r - 1)^{n-1}$ to vanish for $n \geq 1$, we derive:

\begin{widetext}
\begin{equation}\label{equ_Linear_equation_horizonexpansion}
\mathbb{M}\cdot\varphi=
\begin{pmatrix}
M_{11} & 2 (2\pi T_h - i\omega) & 0 & \cdots & 0 \\
M_{21} & M_{22} & 4 (4\pi T_h - i\omega) & \cdots & 0 \\
\vdots & \vdots & \vdots & \ddots & \vdots \\
M_{n1} & M_{n2} & M_{n3} & \cdots & 2n (2n\pi T_h - i\omega)
\end{pmatrix}
\begin{pmatrix}
\varphi_0 \\ \varphi_1 \\ \vdots \\ \varphi_{n}
\end{pmatrix}=
\begin{pmatrix}
0 \\ 0 \\ \vdots
\end{pmatrix}.
\end{equation}
\end{widetext}
The components $M_{ij}$ depend algebraically on $g_{vv_m}$ and $g_{vr_{m-1}}$ from \eqref{equ_general_metric_d}. Once they are specified, the values of $\omega_{n}$ can be obtained by vanishing the last column of $\mathbb{M}$, yielding \mbox{$\omega_{n}=-i2 n\pi T_h$}, while $\mu_n$ can be determined by solving the determinant equation $\text{Det}(\mathcal{M}^{(n)}(\omega_n, \mu)) = 0$, or more compactly, $\text{Det}(\mathcal{M}^{(n)}(\boldsymbol{\mu})) = 0$, where $\mathcal{M}^{(n)}$ denotes the first $n \times n$ submatrix of $\mathbb{M}$. Under these conditions, $\mathbb{M}\cdot\varphi=0$ admits two independent parameters, $\varphi_0$ and $\varphi_n$, thus confirming two ingoing solutions. For the KG equation \eqref{equ_KGequation_General_d}, the associated $\text{Det}(\mathcal{M}^{(n)}(\boldsymbol{\mu}))=0$ defines a degree-$n$ polynomial in $\mu$, which we can express explicitly as
\begin{equation}\label{equ_mu_order_n_vieta}
V_{n, n}\mu^{n}+V_{n, n-1}\mu^{n-1}+\cdots+V_{n, 1}\mu+V_{n, 0}=0,
\end{equation}
where each $V_{n,m}$ denotes the coefficient of the term $\mu^{m}$ in $\text{Det}(\mathcal{M}^{(n)}(\boldsymbol{\mu}))$. By the fundamental theorem of algebra, this degree-$n$ polynomial admits $n$ roots, denoted $\mu_{n,q}$ for $q = 1, \ldots, n$. Owing to its $S_n$ symmetry, one can construct $n$ elementary symmetric polynomials $E_n(\mu^m)$, where $m = 1, \ldots, n$ indicates the degree \cite{macdonald_1999_symmetric_polynomial}. For example, $E_3(\mu^2) = \mu_{3,1}\mu_{3,2} + \mu_{3,2}\mu_{3,3} + \mu_{3,3}\mu_{3,1}$.

{\it Flipping near-horizon analysis---}To proceed with the bulk reconstruction, we now invert the previous near-horizon analysis by interchanging the roles of metric expansion coefficients $(g_{vv_{n}}, \mkern6mu g_{vr_{n-1}})$ and the pole-skipping points $(\omega_{n},\mkern6mu \mu_{n,q})$ in $\text{Det}(\mathcal{M}^{(n)}(\boldsymbol{\mu})) = 0$. By recognizing the latter as given input and the former as unknowns, we reinterpret the determinant equation $\text{Det}(\mathcal{M}^{(n)}(\boldsymbol{\mu})) = 0$ as a system of $n$ equations for $g_{vv_{n}}$ and $g_{vr_{n-1}}$, denoted by $\text{Det}(\vec{\mathcal{M}}^{(n)}(\textbf{g})) = 0$, where each equation corresponds to a distinct choice of $\mu_{n,q}$.

Yet a more natural and illuminating formulation emerges by leveraging the $S_n$ symmetry and labeling the equations in $\text{Det}(\vec{\mathcal{M}}^{(n)}(\mathbf{g})) = 0$ in terms of $n$ elementary symmetric polynomials $E_n(\mu^m)$. These polynomials can be related to $g_{vv_i}$ and $g_{vr_{i-1}}$ via the polynomial coefficients $V_{n,m}$ in Eq. \eqref{equ_mu_order_n_vieta}. Specifically, by invoking Vieta's formula, we obtain
\begin{equation}\label{equ_elementary_symmetric_polynomial_vieta}
E_{n}(\mu^{m})=\frac{v_{n, n-m}}{v_{n, n}},
\end{equation}
where we define $v_{n,m}=(-1)^{n-m}V_{n,m}$ for convenience. Here, $E_n(\mu^m)$ serve as known parameters extracted from the set of pole-skipping points, while the coefficients $v_{n,m}$ are generally algebraic functions of $g_{vv_{i}}$ and $g_{vv_{i-1}}$. Under this reformulation, the $q^{\text{th}}$ equation in $\text{Det}(\vec{\mathcal{M}}^{(n)}(\textbf{g}))=0$, denoted $\text{Det}(\mathcal{M}^{(n)}_q(\textbf{g}))=0$, corresponds to the identity $E_n(\mu^q)- \frac{v_{n, n-q}}{v_{n, n}} =0$.

{\it Bulk reconstruction---}We can then determine $g_{vv_n}$ and $g_{vr_{n-1}}$ by directly solving $\text{Det}(\vec{\mathcal{M}}^{(n)}(\textbf{g})) = 0$. For $n = 1$, $\text{Det}(\vec{\mathcal{M}}^{(1)}(\textbf{g})) = 0$, derived from the KG equation \eqref{equ_KGequation_General_d}, provides one equation: 
\begin{equation}
    E_{1}(\mu)+\frac{d g_{vv_{1}}}{2 g_{vr_{0}}^{2}}=0.
\end{equation}
A second equation arises by equating the boundary temperature $T_b$, expressed in $\omega_1$ as $T_b = i\frac{\omega_1}{2\pi}$, with the Hawking temperature $T_h = \frac{g_{vv_1}}{4\pi g_{vr_0}}$, leading to
\begin{equation}
    i\frac{\omega_1}{2\pi} = \frac{g_{vv_1}}{4\pi g_{vr_0}}.
\end{equation}
Solving these two equations yields:
\begin{equation}\label{equ_sol_gvv1_gvr0_d}
g_{vv_1}=\frac{2 d \omega_1^2}{E_{1}(\mu)}, \quad g_{vr_0}=-\frac{i d \omega_1}{E_{1}(\mu)},
\end{equation}
where both $g_{vv_1}$ and $g_{vr_0}$ are expressed entirely in terms of the pole-skipping data: $\omega_1$ and $E_{1}(\mu)$ (i.e., $\mu_{1,1}$).

We proceed by considering $n=2$. Upon substituting the solutions for $g_{vv_{1}}$ and $g_{vr_{0}}$ from Eq. \eqref{equ_sol_gvv1_gvr0_d} into $\text{Det}(\vec{\mathcal{M}}^{(2)}(\textbf{g}))=0$, we obtain expressions for $g_{vv_{2}}$ and $g_{vr_{1}}$:
\begin{equation}\label{equ_sol_gvv2_gvr1_d}
\begin{aligned}
&g_{vv_{2}}=\frac{d^2 \omega_1^2 E_{2}(\mu^2)}{4 E_{1}(\mu)^3} + \frac{2 d^2 \omega_1^2}{E_{1}(\mu)} - \frac{d^2 \omega_1^2 E_{2}(\mu)}{E_{1}(\mu)^2} + \frac{3 d \omega_1^2}{E_{1}(\mu)},\\ 
&g_{vr_{1}}=\frac{i d^2 \omega_1 E_{2}(\mu)}{2 E_{1}(\mu)^2} - \frac{i d^2 \omega_1}{E_{1}(\mu)} - \frac{i d^2 \omega_1 E_{2}(\mu^2)}{4 E_{1}(\mu)^3} - \frac{i d \omega_1}{E_{1}(\mu)}.\\
\end{aligned}
\end{equation}

For $n > 2$, the system $\text{Det}(\vec{\mathcal{M}}^{(n)}(\textbf{g})) = 0$ becomes overdetermined, consisting of $n$ equations but only two variables: $g_{vv_{n}}$ and $g_{vr_{n-1}}$. However, as proved in \cite{Lu_2024_Longer_paper}, upon incorporating all the solutions for $g_{vv_{m}}$ and $g_{vr_{m-1}}$ with $m<n$ into $\text{Det}(\vec{\mathcal{M}}^{(n)}(\textbf{g}))=0$, the first $n-2$ equations become independent of $g_{vv_{n}}$ and $g_{vr_{n-1}}$. The last two equations: $\text{Det}(\mathcal{M}^{(n)}_{n-1}(\textbf{g}))=0$ and \mbox{$\text{Det}(\mathcal{M}^{(n)}_{n}(\textbf{g}))=0$}, corresponding respectively to
\begin{equation}\label{equ_last_two_equations_vieta}
E_n (\mu^{n-1})- \frac{v_{n, 1}}{v_{n, n}}=0, \quad E_n(\mu^n)- \frac{v_{n, 0}}{v_{n, n}}=0,
\end{equation}
are both \textbf{linear} in $g_{vv_n}$ and $g_{vr_{n-1}}$. Moreover, they are linearly independent, as each involves an elementary symmetric polynomial of a different degree. We thus conclude that for all $n \geq 2$, the pair of equations $\text{Det}(\mathcal{M}^{(n)}_{n-1}(\textbf{g})) = 0$ and $\text{Det}(\mathcal{M}^{(n)}_{n}(\textbf{g})) = 0$ uniquely determines $g_{vv_n}$ and $g_{vr_{n-1}}$, provided a solution exists. They are fully expressible in terms of the pole-skipping data: $\omega_1$, $E_m(\mu^{m-1})$, and $E_m(\mu^{m})$ for all $m \leq n$.

The explicit forms of $g_{vv_{n}}$ and $g_{vr_{n-1}}$ for $n > 2$ are too lengthy to display. Nevertheless, once these coefficients are reconstructed to sufficiently large $n$, $g_{vr}(r)$ and $g_{vv}(r)$ can be approximated to arbitrary precision within the convergence disk of the near-horizon series. The radius of convergence is set by the distance from the horizon at $r = 1$ to the nearest singularities in the complexified $r$-plane. Intriguingly, this implies that the black hole interior is reconstructible, provided no singularities arise before reaching the central singularity at $r = 0$. Similarly, applying the coordinate transformation $z = 1/r$, which places the boundary at $z = 0$, allows reconstruction of the exterior geometry under the analogous assumption. However, for metrics with factors like $1/(1+r^2)$, singularities at $r = \pm i$ limit the convergence domain to a disk of radius $\sqrt{2}$ centered at $r=1$; reconstruction is then valid only within this region.

Moreover, as detailed in \cite{Lu_2024_Longer_paper}, our reconstruction method applies to any massive KG-type equation \mbox{$(\nabla^{2} + m^2)\phi = 0$}, regardless of the specific field type. We adopt a massless scalar field here merely for simplicity. Nevertheless, the massless KG equation \eqref{equ_KGequation_General_d} already captures a broad scenarios beyond the probe limit---including the linearized Einstein equations for tensor perturbations, classified by their transformation properties under $\text{SO}(d-1)$ \cite{Kovtun_2005_QNM_holography}.


{\it Reinterpret Einstein equation---}The analytical reconstruction of the bulk metric immediately implies that the corresponding pure gravitational field equations can be explicitly reformulated in terms of boundary pole-skipping data. The simplest and most important example is the vacuum Einstein equation with a negative cosmological constant:
\begin{equation}\label{equ_vacuum_Einstein_equation}
E_{\mu\nu} \equiv R_{\mu\nu}-\frac{1}{2}g_{\mu\nu}R+\Lambda g_{\mu\nu}=0,
\end{equation}
where $\Lambda = -\frac{d(d+1)}{2}$, with the AdS radius set to unity. Expanding near the horizon up to zeroth order yields three nontrivial components: $E_{vr_{0}} = 0$, $E_{rr_{0}} = 0$, and $E_{xx_{0}} = 0$. Upon substituting the reconstructed metric components from Eqs. \eqref{equ_sol_gvv1_gvr0_d} and \eqref{equ_sol_gvv2_gvr1_d}, these equations can be rewritten entirely in terms of pole-skipping data:
\begin{equation}\label{equ_Evr0_Err0_Exx0_PS}
\begin{aligned}
&E_{vr_{0}}=\frac{d \omega_1}{E_{1}(\mu)} \left (2 E_{1}(\mu) + d^2 + d\right)=0,\\
&E_{rr_{0}}=d \left (4 (d+1) E_{1}(\mu) - 2 d E_{2}(\mu) + \frac{d E_{2}(\mu^2)}{E_{1}(\mu)}\right)=0,\\
&E_{xx_{0}}=6 E_{1}(\mu)-E_{2}(\mu) + d^2 + d =0.\\
\end{aligned}
\end{equation}
This pole-skipping reformulation of the Einstein equations extends to higher orders in the near-horizon expansion. As shown in \cite{Lu_2024_Longer_paper}, the Einstein equation components at order $n$ can be expressed entirely in terms of $\omega_1$, $E_m(\mu^{m-1})$, and $E_m(\mu^{m})$ for all $m \leq n+2$.

{\it $\mu$-polynomial constraints---}Although only the last two equations in $\text{Det}(\vec{\mathcal{M}}^{(n)}(\textbf{g})) = 0$ are directly used in our reconstruction method and reinterpretation of the Einstein equation, we find that the remaining $n - 2$ equations yield $n - 2$ independent polynomial constraints on $\mu_{n,q}$, with degrees $1, 2, \ldots, n - 2$, for any $n > 2$. To explore this further, we first examine $\text{Det}(\mathcal{M}^{(3)}_{1}(\textbf{g})) = 0$, which takes the explicit form: 

\begin{equation}\label{equ_linear_mu_constrain_origin}
E_{3}(\mu)+\frac{3 d g_{vv_1}}{2 g_{vr_0}^2}+\frac{8 g_{vv_1}}{g_{vr_0}^2}+\frac{4 g_{vr_1} g_{vv_1}}{g_{vr_0}^3}-\frac{8 g_{vv_2}}{g_{vr_0}^2}=0.
\end{equation}
This equation does not involve $g_{vv_3}$ or $g_{vr_2}$ and thus does not contribute to their reconstruction, as previously stated. After substituting the solutions for $g_{vv_1}$, $g_{vr_0}$ from Eq. \eqref{equ_sol_gvv1_gvr0_d} and $g_{vv_2}$, $g_{vr_1}$ from Eq. \eqref{equ_sol_gvv2_gvr1_d}, Eq. \eqref{equ_linear_mu_constrain_origin} simplifies to a homogeneous polynomial identity in $\mu$:
\begin{equation}\label{equ_linear_mu_constrain_intermediate}
E_{3}(\mu)-4E_{2}(\mu)+5E_{1}(\mu)=0,
\end{equation}
or more intuitively,
\begin{equation}\label{equ_linear_mu_constrain_3}
\mu_{3,1} + \mu_{3,2} + \mu_{3,3} - 4(\mu_{2,1}+\mu_{2,2})+5\mu_{1,1}=0.
\end{equation}
This identity implies that the three variables $\mu_{3,q}$ are not all independent: e.g., $\mu_{3,3}$ can be determined from the other five $\mu_{n,q}$ with $n \leq 3$ using identity \eqref{equ_linear_mu_constrain_3}.

At $n = 4$, we obtain similar homogeneous polynomial identities from the equations $\text{Det}(\mathcal{M}^{(4)}_{1}(\mathbf{g})) = 0$ and $\text{Det}(\mathcal{M}^{(4)}_{2}(\mathbf{g})) = 0$ after incorporating the previously solved values of $g_{vv_n}$ and $g_{vr_{n-1}}$ for $n = 1, 2, 3$. These two equations reduce to:
\begin{equation}\label{equ_P4_mu_linear}
E_{4}(\mu)-10E_{2}(\mu)+16E_{1}(\mu)=0,
\end{equation}
and
\begin{equation}\label{equ_P4_mu_quadratic_invariant}
\begin{aligned}
&E_{4}(\mu^2)-6 E_{3}(\mu^2)+14 E_{2}(\mu^2)-9 E_{2}(\mu)^2\\
&+40 E_{2}(\mu) E_{1}(\mu)-46 E_{1}(\mu)^2=0,\\
\end{aligned}
\end{equation}
respectively. These two polynomial identities reduce the number of independent variables among $\mu_{4,q}$ from four to two. We then refer to such identities as $\mu$-polynomial constraints. 

For larger $n$, a similar pattern emerges, as detailed in \cite{Lu_2024_Longer_paper}: by substituting the solutions for $g_{vv_m}$ and $g_{vr_{m-1}}$ obtained from $\text{Det}(\vec{\mathcal{M}}^{(m)}(\mathbf{g})) = 0$ for all $m < n$ into $\text{Det}(\vec{\mathcal{M}}^{(n)}(\mathbf{g})) = 0$, each equation $\text{Det}(\mathcal{M}^{(n)}_{m}(\mathbf{g})) = 0$ with $m = 1, 2, \ldots, n - 2$ reduces to a $\mu$-polynomial constraint of degree $m$. We denote these as $P_n(\mu^m) = 0$, where $n$ is the expansion order and $m$ is the degree of the polynomial. For example, the constraint in Eq. \eqref{equ_linear_mu_constrain_3} is denoted as $P_3(\mu) = 0$.

Generally, at expansion order $n$, the set of equations $\text{Det}(\mathcal{M}^{(n)}_{1}(\textbf{g}))=0$, $\cdots$, $\text{Det}(\mathcal{M}^{(n)}_{n-2}(\textbf{g}))=0$ yields $n-2$ $\mu$-polynomial constraints: $P_n(\mu)=0$, $\cdots$, $P_{n}(\mu^{n-2})=0$, which reduce the number of independent $\mu_{n,q}$ from $n$ to $2$, consistent with the number of the bulk variables $g_{vv_n}$ and $g_{vr_{n-1}}$. Although these variables are expressed through $E_n(\mu^m)$ involving all $\mu_{n,q}$, these constraints ensure that specifying any two of $\mu_{n,q}$ for each $n > 2$ suffices; the remaining $\mu_{n,q}$ are fixed by the $\mu$-polynomial constraints.

As demonstrated in \cite{Lu_2024_Longer_paper}, these constraints remain valid in the context of the master equation \mbox{$(\nabla^2 + V(r))\Phi(r) = 0$}, where $\nabla$ is the covariant derivative with respect to the background metric \eqref{equ_general_metric_d}, the master field $\Phi(r)$ may encompass perturbations from various sectors (scalar, vector, and tensor) \cite{kodama_2003_Master_fields_0, kodama_2004_master_fields_1, jansen_2019_Master_fields_KG_type}, and $V(r)$ denotes the corresponding potential. The only requirement is that the associated $\text{Det}(\mathcal{M}^{(n)}(\boldsymbol{\mu}))$ remains a degree-$n$ polynomial in $\mu$. 

Importantly, the validity of these constraints does not depend on the spacetime dimension or the asymptotic structure of the background geometry, since the metric \eqref{equ_general_metric_d} imposes no such restrictions. This universality is illustrated through various examples in \cite{Lu_2024_Longer_paper}, including the KG equation for a probe scalar in four-dimensional Lifshitz black holes and the longitudinal master equation for $U(1)$ gauge field perturbations in $d+2$-dimensional AdS black holes, among others.

We conclude by presenting the general formula for $P_n(\mu)$:
\begin{equation}\label{equ_Pn_mu_linear_explicit} \\
E_{n}(\mu) - \frac{1}{6}\frac{(n+1)!}{(n-2)!}E_{2}(\mu)+\frac{1}{3}\frac{(n+2)!!}{(n-4)!!}E_{1}(\mu)=0.
\end{equation}
This formula reproduces $P_3(\mu)=0$ \eqref{equ_linear_mu_constrain_3} and $P_4(\mu) = 0$ \eqref{equ_P4_mu_linear} when setting $n = 3$ and $n = 4$, respectively \footnote{Factorial expressions should be simplified before substituting specific values of $n$.}. A rigorous proof of the general expression \eqref{equ_Pn_mu_linear_explicit} is provided in \cite{Lu_2024_Longer_paper}.

{\it Discussion---}We develop an analytical boundary-to-bulk map that reconstructs the near-horizon geometry from boundary pole-skipping points by solving a system of linear equations. Provided the metric remains analytic, this reconstruction extends across both the interior and exterior of the black hole, bounded only by the nearest singularity.

\begin{figure}[h!] 
\centering 
\includegraphics[width=0.49\textwidth]{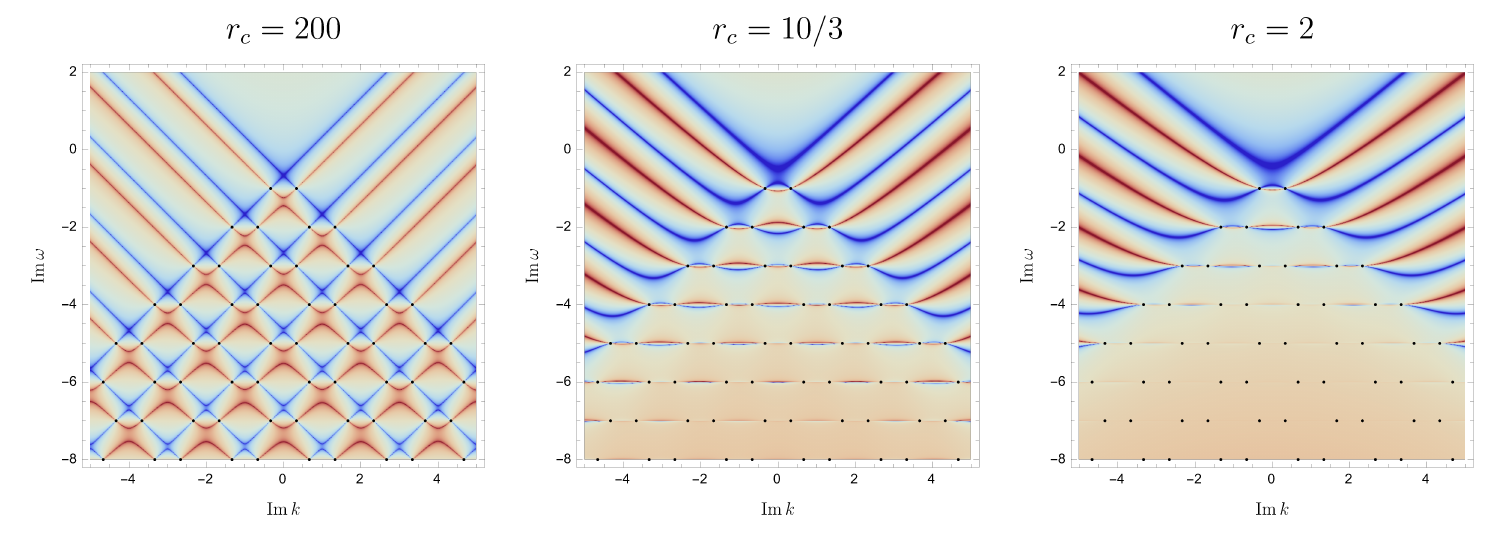} 
\caption{\label{fig_TTbar_GR} 
Heatmaps of the Green's function in deformed CFTs dual to BTZ black holes with a finite radial cutoff. Red and blue lines denote the poles and zeros respectively, while black dots mark the pole-skipping points. From left to right, the location of the radial cutoff surface $r_c$ approaches the horizon, making higher-order poles and zeros increasingly difficult to distinguish.} 
\end{figure}

With this map in hand, a natural question arises: how is it affected when the boundary field theory is deformed? A useful setting to explore this is the $T\bar{T}$ deformation \cite{cavaglia_2016_TbarT-deformation_original_2a, smirnov_2017_TbarT-Deformation_original_1}, which holographically corresponds to introducing a finite radial cutoff in the bulk \cite{mcgough_2018_TbarT-deformation_Holographya}. As illustrated in Fig. \ref{fig_TTbar_GR}, we observe that as more of the near-boundary region of the bulk spacetime is removed, the poles and zeros of Green's function begin to blur together, making the higher-order pole-skipping points increasingly difficult to identify in practice.

\begin{figure}[h!]
\centering
\includegraphics[width=0.43\textwidth]{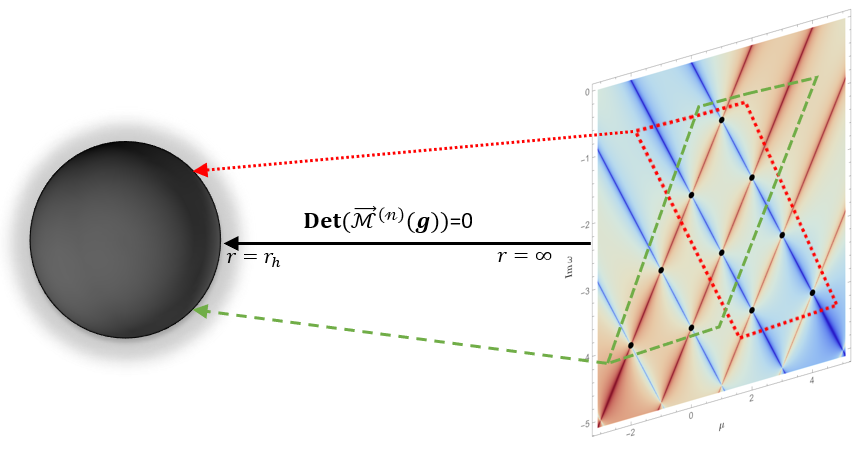}
\caption{\label{fig_explicit_map} The black hole geometry can be reconstructed from different subsets of boundary pole-skipping points (represented by black dots) by solving the linear system $\text{Det}(\vec{\mathcal{M}}^{(n)}(\textbf{g})) = 0$. The green and red quadrilaterals enclose portions of two such subsets.}
\end{figure}

Moreover, our method reveals that the bulk metric is redundantly encoded in the boundary pole-skipping points, as illustrated schematically in Fig. \ref{fig_explicit_map}. This redundancy arises from a set of $\mu$-polynomial constraints, such as $P_{n}(\mu)$  \eqref{equ_Pn_mu_linear_explicit}. Upon imposing these constraints, the remaining independent pole-skipping points precisely match the number of bulk degrees of freedom, namely $g_{vv_{n}}$ and $g_{vr_{n-1}}$. 

Previously, Ref. \cite{grozdanov_2023_Pole_skipping_reconstruct_spectrum} argued that the infinite poles-kipping points lying along a single hydrodynamic mode, expanded in large spacetime dimensions, suffice to reconstruct the quasinormal spectrum and related Green's function. Suppose that the geometry uniquely determines the Green's function. Our method immediately implies that the latter can be recovered from the pole-skipping points, for any spacetime dimension. This validates the results in \cite{grozdanov_2023_Pole_skipping_reconstruct_spectrum}, albeit via a different route. Furthermore, the high redundancy in the pole-skipping data sheds some light on the interesting question raised in \cite{grozdanov_2023_Pole_skipping_reconstruct_spectrum}:  What is the minimum information required to determine the spectrum of a correlation function?

We emphasize that the implication of $\mu$-polynomial constraints is not exclusive to holographic theories. Without relying on the existence of a dual field theory, these constraints can be derived under a general static, planar-symmetric black hole background, along with a class of master equations of the form $(\nabla^2 + V(r))\Phi(r) = 0$, whose pole-skipping structure aligns with the KG equation \eqref{equ_KGequation_General_d}. Generalizing these constraints to spherical or hyperbolic black holes is straightforward. Turning back to holography, it would be intriguing to investigate the origin of these constraints from the field theory perspective and to explore their applicability to the energy density Green's functions that exhibit chaotic pole-skipping points at the upper-half frequency plane.

Finally, owing to its analyticity and simplicity, our reconstruction method offers a compelling framework that could drive future experiments exploring the emergence of spacetime, e.g., the search for ``spacetime-emergent materials" \cite{koji_2022_space_emergent_material, koji_2025_ML_Green_function_reconstruct}. However, pole-skipping points are complex in most holographic models and therefore not directly measurable. A potential solution is analytic continuation, inferring complex-valued pole-skipping data from real experimental measurements or Euclidean numerical simulations \cite{Gattringer_2010_Lattice_QCD_introduction_2, Foulkes_2001_QMC_introduction}. Although this constitutes an ill-posed inverse problem, various numerical techniques \cite{Jarrell_1996_analytic_continuation_introduction, Shao_2022_analytic_continuation_introduction_2} have been developed and continue to improve, e.g., see the barycentric rational method \cite{Huang_2025_barycentric_analytic_continuation}.

In certain systems, such as AdS-solitons \cite{Natsuume_2023_pole-skipping_non-black_hole} and holographic lattices \cite{balm_2020_Holography_fermion_lattice_potential}, pole-skipping points can be real-valued. Similar phenomena involving real zero-pole interference also occur in condensed matter \cite{sakai_2009_Mott_Simulator_poles_meet_zeros_simulation, sakai_2010_Mott_Simulator_poles_meet_zeros_simulation_2, Misawa_2022_PS_detect_topological_insulator}. A primary challenge to achieving the pole-skipping here would be that conventional spectroscopic methods resolve poles but are insensitive to zeros. Nevertheless, specific schemes such as cotunneling experiments have been proposed to directly probe the zeros \cite{Lehmann_2025_probe_zeros_experiment}. Looking ahead, quantum processors may eventually enable the direct extraction of pole-skipping data in many-body systems \cite{kokcu_2024_Quantum_simulation_bosonfermion_correlation}.


{\it Acknowledgments---}We would like to thank Navid Abbasi, Xian-Hui Ge, Song He, Yan Liu and Zhuo-Yu Xian for helpful discussions. A special thanks goes to Sašo Grozdanov and Mile Vrbica for their frequent and insightful discussions, as well as their valuable feedback on the draft of this Letter. SFW is supported by NSFC grants No.12275166 and No.12311540141.




%

\end{document}